\documentclass{mn2e}
\usepackage{graphicx}

\title[On the reconstruction of a magnetosphere of pulsars nearby the
light cylinder surface]{On the reconstruction of a magnetosphere
of pulsars nearby the light cylinder surface}
\author[Z. Osmanov, G. Dalakishvili and G. Machabeli]{Z. Osmanov$^{1}$\thanks{E-mail:
z.osmanov@astro-ge.org (ZO); giorgidalakishvili@hotmail.com (GD);
g.machabeli@astro-ge.org (GM) } G. Dalakishvili$^{2}$ and G.
Machabeli$^{1,2}$ \\ $^{1}$Georgian National Astrophysical
Observatory, Kazbegi str. $2^a$, Tbilisi, 38006, Georgia\\
$^{2}$Department of Physics, Faculty of Exact and Natural
Sciences, Tbilisi State University, Chavchavadze ave $1$, Tbilisi,
0128, Georgia }

\begin{document}

\date{Accepted 1988 December 15. Received 1988 December 14; in original form 1988 October 11}

\pagerange{\pageref{firstpage}--\pageref{lastpage}} \pubyear{2002}

\maketitle

\label{firstpage}

\begin{abstract}
A mechanism of generation of a toroidal component of large scale
magnetic field, leading to the reconstruction of the pulsar
magnetospheres is presented. In order to understand twisting of
magnetic field lines, we investigate kinematics of a plasma stream
rotating in the pulsar magnetosphere. Studying an exact set of
equations describing the behavior of relativistic plasma flows,
the increment of the curvature drift instability is derived, and
estimated for $1s$ pulsars. It is shown that a new parametric
mechanism is very efficient and can explain rotation energy
pumping in the pulsar magnetospheres.\end{abstract}


\section{Introduction}

The aim of the present work is to investigate  generation of a
toroidal component of the magnetic field nearby the light cylinder
surface (LCS) (a hypothetical surface, where the linear velocity
of rotation equals the speed of light).

The work we consider in this paper is closely related to the
pulsar wind problem. Studying the magnetic field of the Crab
nebula, Piddington (1953) was first who has suggested the presence
of a central object in the nebula, with frozen magnetic field
inside. It has been supposed that rotation of the central body
provokes generation of the toroidal component of magnetic field.
Further investigations have shown that this kind of magnetic field
characterizes magnetized star winds \cite{weber}. These results
have been generalized for relativistic flows in a region close to
the LCS: \cite{michel,ken1,ken2,beg}. Despite success of developed
models, they encounter a number of difficulties, when one attempts
to extrapolate the wind back to the source: the pulsar
magnetosphere. For large distances the wind is specified in the
approximation: $\sigma\equiv B^2/(4\pi mc^2n\gamma)\ll 1$, where
$B$ is the magnetic field induction, $m$ and $n$- the electron
mass and density respectively and $\gamma$ the Lorentz factor of
relativistic electrons. In this case, change of magnetic field's
configuration is defined only by plasma motion. This circumstance
simplifies a possibility of analytical consideration of a plasma.
But in the pulsar magnetospheres, a situation is opposite, the
energy density of magnetic field exceeds by many orders of
magnitude the energy density of the plasma $\sigma\gg 1$,
therefore a need of consideration of this specific case is
essential. Close to the light cylinder area the magnetic field
drags behind itself the rotating electron-positron plasma and the
question which arises is: how the magnetosphere is reconstructed
nearby the light cylinder surface? It is obvious that close to
this region, rigid rotation is impossible and consequently
magnetic field lines must deviate, lagging behind the rotation of
the pulsar. Implementing special MHD codes in a series of works
\cite{michel1,michel2,smith} pulsar wind physics has been
numerically studied and improved by Spitkovski \& Arons (2002) and
Spitkovski (2003) where plasma dynamics in 3D was presented and it
has been shown that the flow goes through the LCS into the wind
zone. In these papers a principal assumption is the current
generated by the electric drift: $\overrightarrow{V_E} =
c\overrightarrow{E}\times\overrightarrow{B}/B^2$ \cite{bland}.
Obviously for a plasma composed of equal numbers of positive and
negative charges, the current is not generated (the electric drift
does not "feel" charges), although for the pulsar plasma a primary
electron beam is composed of only electrons and therefore the
electric drift generates the current, leading to creation of
electromagnetic fields.

 In \cite{r03} a particle moving along a curved rotating channel has been considered  and it was
shown that for a certain shape of curved trajectories one may
avoid the light cylinder problem. Therefore one has to understand
what is a mechanism responsible for the process of twisting of
field lines when the condition $\sigma\gg 1$ is satisfied.

According to observations it is clear that the energy of emission
is very high. An observed pulsar luminosity lies in the range:
$[10^{31}-10^{38}]erg/s$ \cite{catalog}, on the other hand the
only source of pulsar radiation can be rotational energy
$I\omega^2/2$, where $I$ is moment of inertia of the pulsar, and
$\omega$ - the angular velocity of rotation. As observations show
the spin down luminosity is of the same order of magnitude as the
radiation luminosity, therefore it is reasonable to suppose that
all pulsars emit due to rotation energy decrease \cite{stur}. The
problem concerns the question: how the rotation energy is
transformed into pulsar radiation. According to standard models,
due to electric field, the charged particles uproot from a surface
of the neutron star and accelerate by the electric force which
results in the radiation process. The origin of this emission is
supposed to be in the magnetosphere of pulsars. These models
introduce a vacuum gap, inside of which the particles experience
strong electric field and accelerate. But the problem arises
concerning the gap size which turns out to be not enough for
energy gain of charged particles \cite{rud}.

In order to resolve this problem and enlarge the gap size (which
will provide increase of an acceleration length scale) many
attempts have been done, applying different approaches:
\cite{aar,mus,rud}, but no approach was able to get the efficient
acceleration enough for producing observed radiation.

A new mechanism of acceleration has been introduced in \cite{mr}
where a bead moving inside a straight rigidly rotating pipe has
been studied. It was shown that the centrifugal force can be very
efficient and if one applies this method for the pulsar
magnetospheres it will provide high Lorentz factors of particles.
Therefore the amount of energy contained within the $e^{+}e^{-}$
plasma is very high. If one finds mechanisms for the conversion of
at least a small fraction of this energy into the variety of waves
or instabilities - one might witness a number of well-pronounced
and bona fide observational signatures in the pulsar radiation
theory. In \cite{mr} it has been found that the radial component
of velocity for relativistic particles behaves in time as
$c\cos(\Omega t)$ ($c$ is the speed of light), which gives a
possibility of parametric energy pumping from the mean flow into
instabilities (see \cite{incr}). In \cite{incr} the $e^{+}e^{-}$
plasma has been studied and the increment of an instability of the
Lengmuire waves was estimated. It has been demonstrated that the
centrifugal acceleration might have been efficient enough for the
observed spin down luminosity. We have shown that the linear stage
was so efficient that it was very short in time, and
nonlinearities were turned in soon.

In the present paper we generalize the previous work and study the
parametric mechanism of the curvature drift instability driven by
the centrifugal acceleration. We consider a two component plasma:
a) the basic plasma flow (bulk flow) with the concentration
$n_{pl}$ and the Lorentz factor $\gamma_{pl}$ and b) the beam
component with the concentration $n_{b}$ and the Lorentz factor
$\gamma_{b}$. It is known that in the pulsar magnetosphere the
drift velocity is to be important for plasma dynamics. The drift
velocity may influence processes in the plasma and especially may
affect an evolution of instabilities. Unlike \cite{spit1,spit2}
where the processes are considered nearby the pulsar surface, in
the present paper we investigate instabilities close to the light
cylinder area, where effects of centrifugal acceleration should be
extremely efficient. In \cite{spit2} it has been noted that the
structure of pulsar magnetospheres could not be solved
analytically, whereas in the present paper, we show that an
initial stage of the reconstruction process of magnetospheres can
be considered analytically, starting by appropriate initial
conditions. Another difference is that in our model we study a
plasma, which is bound by rigidly rotating straight magnetic field
and the force free condition applied in \cite{spit1,spit2} is not
valid, because as it is shown in \cite{shafo} the force free
condition can be provided only if the magnetic field has a
configuration similar to the one of a differentially rotating
Couette flow. The principally different assumption in the present
paper is that instead of considering the electric drift, we study
the curvature drift investigating the possibility of generation of
the toroidal component $B_r$, which is a key step in understanding
the reconstruction of the pulsar magnetosphere nearby the LCS.

The work is organized as follows. In \S\ref{sec:disper} we derive
the dispersion relation, in \S\ref{sec:discus} the corresponding
results are present and in \S\ref{sec:summary} we summarize the
results.

\begin{figure}
 \par\noindent
 {\begin{minipage}[t]{1.\linewidth}
 \includegraphics[width=\textwidth] {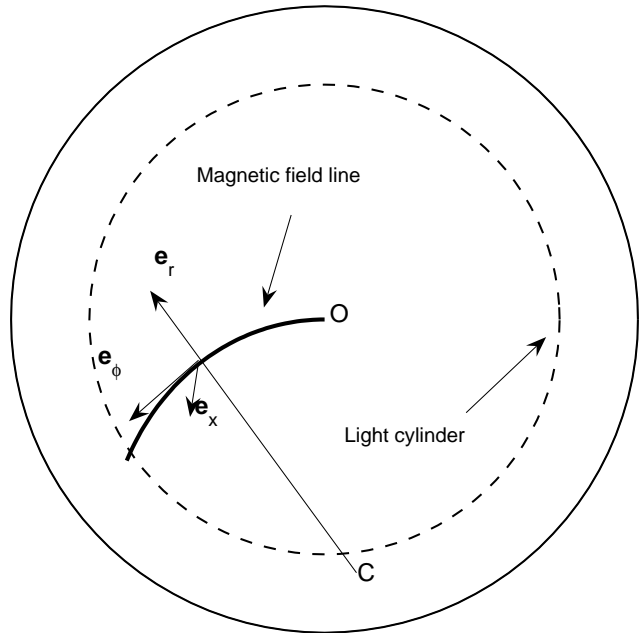}
 \end{minipage}
 }
 \caption[ ] {Here we show geometry in which we consider our system of equations.
 By ${\bf e_{\phi}}$, ${\bf e_{r}}$ and ${\bf e_{x}}$ unit vectors are denoted, note that
 ${\bf e_{x}}\perp {\bf e_{r,\phi}}$. $C$ is the curvature center.}\label{geom}
 \end{figure}
\section{Theory} \label{sec:disper}
%
%
%
Throughout the work it is supposed, that magnetic field lines are
almost straight and due to the frozen-in condition the plasma
particles follow the magnetic lines and accelerate. Geometry in
which we consider the problem is shown in Fig. \ref{geom}
\cite{lmb}.

Our system is governed by the Euler equation \cite{incr}:

\begin{equation}
\label{eul} \frac{\partial{\bf p_i}}{\partial t}+({\bf
v_i\nabla)p_i}=
-\gamma\alpha{\bf\nabla}\alpha+\frac{e}{m}\left(\bf E+ \bf
v_i\times\bf B\right),
\end{equation}
$$i = pl,b, $$ the continuity equation:
\begin{equation}
\label{cont} \frac{\partial n_i}{\partial t}+{\bf \nabla}(n_i{\bf
v_i})=0,
\end{equation}
and the induction equation, which closes the system:
\begin{equation}
\label{ind} {\bf \nabla\times B} = \frac{1}{c}\frac{\partial {\bf
E}}{\partial t}+\frac{4\pi}{c}\sum_{i=pl,b}{\bf J_i},
\end{equation}
where ${\bf J_i}$ ($i = pl,b$) is  the current of plasma and beam
components.

We start our analysis by introducing small deviations around the
equilibrium state:

\begin{equation}
\label{expansion} \Psi\approx \Psi^0 + \Psi^1,
\end{equation}
where $\Psi = (n,{\bf v},{\bf p},{\bf E},{\bf B})$.

Since we are interested in the generation of the toroidal
component of magnetic field, it is interesting to study the
curvature drift wave (when $\omega\sim ku_x$), because it is
characterized by the following conditions: $B_r\gg B_{\phi}$,
$E_{\phi}\gg E_r$ \cite{kazb} that show an importance of $B_r$ on
the one hand and closes the system by the second condition on the
other hand.

We consider the equilibrium state with a drift velocity along the
$x$-axis

\begin{equation}
\label{drift} u_{0x} = \frac{\gamma v_{0\phi}^2}{\omega_B R_B}.
\end{equation}
where $\gamma$ is the Lorentz factor, $\omega_B = eB_0/(m_ec)$ and
$R_B$ is the curvature radius of the magnetic field lines ($e$ and
$m$ are the charge and mass of electron and $B_0$-the magnetic
induction). Along $\phi$, due to the centrifugal acceleration one
has a relativistic flow with the velocity \cite{mr}:

\begin{equation}
\label{vfi} v_{0\phi} = c\cos(\Omega t).
\end{equation}

If one expresses the perturbation of physical quantities by
following:

\begin{equation}
\label{pert} \Psi^1(t,{\bf r})\propto\Psi^1(t)
\exp\left[i\left({\bf kr} \right)\right] \,,
\end{equation}
then considering only $x$ components of the Euler and induction
equation, it is easy to show that for curvature drift waves,
propagating perpendicular to magnetic filed lines ($k_{\phi}\ll
k_x$), Eqs. (\ref{eul},\ref{cont},\ref{ind}) can be reduced into
the form:

\begin{equation}
\label{eulp} \frac{\partial p^1_{ix}}{\partial
t}-i(k_xu_{0x}+k_{\phi}u_{0\phi})p^1_{ix}=
\frac{e}{c}v_{0\phi}B^1_r,
\end{equation}
\begin{equation}
\label{contp} \frac{\partial n^1_{i}}{\partial
t}-i(k_xu_{0x}+k_{\phi}u_{0\phi})n^1_{i}= ik_xn_{i0}v^1_x,
\end{equation}
\begin{equation}
\label{indp} -ik_{\phi}cB^1_r = 4\pi
e\sum_{i=pl,b}(n_{i0}v^1_{ix}+n^1_{i}v_{i0x}).
\end{equation}

In Eq. (\ref{eulp}) we have used an approximate expression of
velocity along the $r$-axis: $v^1_r\approx cE^1_x/B_{0\phi}$. If
we choose $p^1_{ix}$ and $n^1_{i}$ to have the form:

\begin{equation}
\label{anzp} v^1_{ix}\equiv V_{ix}e^{i{\bf kA_i}(t)},
\end{equation}
\begin{equation}
\label{anzn} n^1_{i}\equiv N_ie^{i{\bf kA_i}(t)},
\end{equation}
where
\begin{equation}
\label{Ax} A_x(t) = \frac{U_{ix}t}{2} +
\frac{U_{ix}}{4\Omega}\sin(2\Omega t),
\end{equation}
\begin{equation}
\label{Af} A_{\phi}(t) = \frac{c}{\Omega}\sin(\Omega t),
\end{equation}
\begin{equation}
U_{ix} = \frac{c^2\gamma_{i0}}{\omega_BR_B},
\end{equation}
then one obtains from Eqs. (\ref{eulp},\ref{contp}):

\begin{equation}
\label{vx} v^1_{ix} = \frac{e}{m\gamma_{i0}}e^{i{\bf
kA_i}(t)}\int^te^{-i{\bf kA_i}(t')}v_{0\phi}(t')B_r(t')dt',
\end{equation}

$$n^1_{i} = \frac{ien_{0ik_x}}{m\gamma_{i0}}e^{i{\bf
kA_i}(t)}\int^tdt'\int^{t''}e^{-i{\bf
kA_i}(t'')}v_{0\phi}(t'')B_r(t'')dt''.$$
\begin{equation}
\label{n}
\end{equation}

Substituting Eqs. (\ref{vx},\ref{n}) into Eq. (\ref{indp}) it
reduces to the form:

$$ -ik_{\phi}cB^1_r(t)
=\sum_{i=pl,b}\frac{\omega^2_{i}}{\gamma_{i0}}e^{i{\bf
kA_i}(t)}\int^te^{-i{\bf kA_i}(t')}v_{0\phi}(t')B_r(t')dt'+ $$
$$i\sum_{i=pl,b}\frac{\omega^2_{i}}{\gamma_{i0}}k_xu_{0ix}e^{i{\bf
kA_i}(t)}\int^tdt'\int^{t''}e^{-i{\bf
kA_i}(t'')}v_{0\phi}(t'')B_r(t'')dt'',$$

\begin{equation}
\label{ind1}
\end{equation}
where $\omega_i = \sqrt{4\pi n_{i0}e^2/m}$ is the plasma
frequency. In order to solve this equation one has to take the
Fourier time transform. For this reason if one uses the following
identity:
\begin{equation}
\label{bess} e^{\pm ix\sin\Omega t}=\sum_s J_s(x)e^{\pm is\Omega
t},
\end{equation}
one can reduce Eq. (\ref{ind1}):

$$B_r(\omega) =
-\sum_{i=pl,b}\frac{\omega^2_{i}}{2\gamma_{i0}k_{\phi}c}\sum_{\sigma
= \pm 1}\sum_{s,n,l,p}\frac{J_s(g_i)J_n(h)J_l(g_i)J_p(h)}{\omega +
\frac{k_xU_{ix}}{2}+\Omega (2s+n) } \times$$ $$\times
B_r\left(\omega+\Omega
\left(2[s-l]+n-p+\sigma\right)\right)\left[1-\frac{k_xU_{ix}}{\omega
+ \frac{k_xU_{ix}}{2}+\Omega (2s+n)}\right]$$
$$+\sum_{i=pl,b}\frac{\omega^2_{i}k_xU_{ix}}{4\gamma_{i0}k_{\phi}c}\sum_{\sigma,\mu
= \pm
1}\sum_{s,n,l,p}\frac{J_s(g_i)J_n(h)J_l(g_i)J_p(h)}{\left(\omega +
\frac{k_xU_{ix}}{2}+\Omega (2[s+\mu]+n)\right)^2 } \times$$
\begin{equation}
\label{disp} \times B_r\left(\omega+\Omega
\left(2[s-l+\mu]+n-p+\sigma\right)\right),
\end{equation}
where

$$g_i = \frac{k_xU_{ix}}{4\Omega},$$ $$h =
\frac{k_{\phi}c}{\Omega}.$$

\section{Discussion} \label{sec:discus}
%
%
%
%
%
One can see from the dispersion relation that the system is
characterized by two different kinds of resonance, which come from
the first and second terms of the right hand side of Eq. (
\ref{disp}):

\begin{equation}
\label{res1} \omega + \frac{k_xU_{ix}}{2}+\Omega (2s+n)\simeq 0
\end{equation}
and

\begin{equation}
\label{res2} \omega + \frac{k_xU_{ix}}{2}+\Omega
(2[s+\mu]+n)\simeq 0,
\end{equation}
$$s,n = \{0,\pm 1,\pm 2 ...\},$$

$$\mu = \pm 1.$$
\begin{figure}
 \par\noindent
 {\begin{minipage}[t]{1.\linewidth}
 \includegraphics[width=\textwidth] {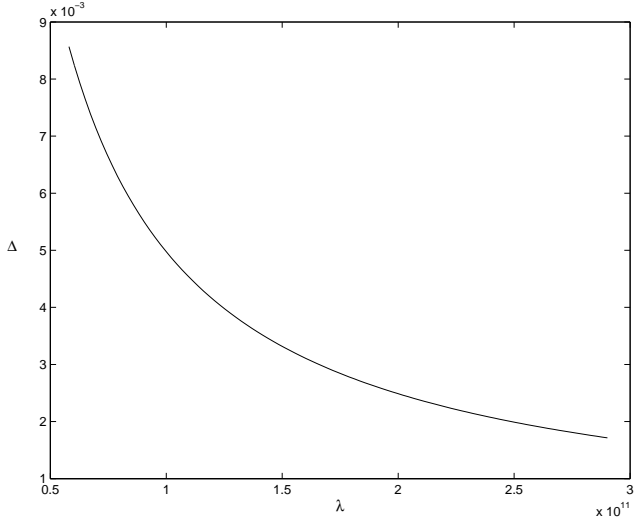}
 \end{minipage}
 }
 \caption[ ] { Dependence of the increment on
$\lambda$ nearby the LCS. The set of parameters is $P = 1s$,
$\gamma_b\sim 10^8$, $\lambda_{\phi} = 1000R_{lc}$. }\label{lam}
 \end{figure}
As we will see later, a resonance frequency from the first
condition Eq. (\ref{res1}) does not influence the corresponding
resonance from the second expression, because if the first term is
valid, the second condition is not satisfied and vice versa, when
the second resonance works, the first one is not valid.

Since we are studying the mechanism of energy pumping from
pulsar's rotation and a natural consequence, that magnetic field
lines must be twisted nearby the light cylinder and the twist
should have a direction opposite to the rotation, one can suppose
that a frequency responsible for this process must be small
compared the angular velocity of the pulsar. On the other hand,
assuming $\lambda\sim 6\times 10^{10} - 3\times 10^{11}cm$ (where
$\lambda\approx\lambda_x$ is the wave length) for {\it second}
pulsars it is straightforward to check that $|k_xU_{ix}/2|\sim
0.02 - 0.1$, which for any values of $s$ and $n$ is much less than
$\Omega (2s+n)$ for $s,n\neq 0$ (see Eq.\ref{res1}). Thus the only
possibility which provides low frequency waves from the first
resonance condition is:

\begin{equation}
\label{filt1} 2s+n = 0.
\end{equation}

Here we have assumed that $R_B\sim R_p\sim 10^6cm$ ($R_p$ is the
pulsar radius). Unlike this case, second resonance condition does
not provide low frequencies (see Eq.(\ref{res2})), because even
for vanishing $s$ and $n$, $\mu$ is not vanishing and hence it
does not contribute in the process of magnetic field line's
twisting.

Let us consider the dispersion relation near the beam resonant
condition expressed by Eq. (\ref{res1}). Then only resonant terms
will be preserved and Eq. (\ref{disp}) will reduce:

$$B_r(\omega_0) \approx
-\frac{\omega^2_{b}k_xU_{bx}}{2\gamma_{b0}k_{\phi}c}\sum_{\sigma =
\pm
1}\sum_{s,l,p}\frac{J_s(g_b)J_{-2s}(h)J_l(g_b)J_p(h)}{\widetilde{\Delta}^2}
\times$$
\begin{equation}
\label{disp1} \times B_r\left(\omega_0+\Omega
\left(-2l-p+\sigma\right)\right),
\end{equation}
where

\begin{equation}
\label{freq} \omega_0\approx -\frac{k_xU_{bx}}{2}
\end{equation}
and the frequency has been expressed by the form:

\begin{equation}
\label{freq1} \omega\equiv\omega_0+i\widetilde{\Delta}.
\end{equation}

Here $\widetilde{\Delta}$'s imaginary part $\Delta\equiv
Im(\widetilde{\Delta})$ is related to the increment of the
instability. Since a dominant term in Eq. (\ref{disp1}) comes from
low frequencies ($\omega_0\ll\Omega$), then the only terms
contributed in a time average will have $p$ equal to $\Omega
(2(s-l)+n+\sigma)$, because all other terms with
$B_r(\omega_0+\Omega q)$ ($q \neq 0$) give zero due to an
oscillative character with very big values of frequencies. Taking
into account this condition, one gets:

$$B_r(\omega_0) \approx
-\frac{\omega^2_{b}k_xU_{bx}}{2\gamma_{b0}k_{\phi}c}\sum_{\sigma =
\pm
1}\sum_{s,l}\frac{J_s(g_b)J_{-2s}(h)J_l(g_b)J_{-2l+\sigma}(h)}{\Delta^2}
\times$$
\begin{equation}
\label{disp2} \times B_r(\omega_0).
\end{equation}

From here one can easily express the increment by following:

\begin{equation}
\label{incr}
\Delta\approx\left[-\frac{\omega^2_{b}k_xU_{bx}}{2\gamma_{b0}k_{\phi}c}
\Sigma_1(g_b,h)\Sigma_2(g_b,h)\right]^{\frac{1}{2}},
\end{equation}
where

\begin{equation}
\label{sig1} \Sigma_1(g_b,h)\equiv \sum_sJ_s(g_b)J_{-2s}(h),
\end{equation}
\begin{equation}
\label{sig2} \Sigma_2(g_b,h)\equiv \sum_{\sigma = \pm 1
}\sum_lJ_l(g_b)J_{-2l+\sigma}(h).
\end{equation}

Strictly speaking $\Sigma_1$ and $\Sigma_2$ are functions of $g_b$
and $h$ and one can  show, that these summations are convergent
(see Appendix).

It is interesting to investigate the increment versus following
physical quantities: $\lambda_{x}$ (for the fixed, and very big in
comparison with $\lambda_{x}$ values of $\lambda_{\phi}$). On the
other hand one has to compare results with an observational
evidence. As we have already mentioned the only source that may
provide energy for radiation is the slowdown of the pulsar: $\dot
W\approx I\Omega\dot \Omega$, here $I$ is moment of inertia of the
pulsar. The rate of rotation energy loss can be estimated by
following ratio: $\dot W/W\simeq2\dot\Omega/\Omega=2\dot P/P$,
where $P$ is rotation period of the pulsar. The given ratio is
different for different pulsars and ranges from
$10^{-11}\sec^{-1}$ (PSR 0531) to $10^{-18}\sec^{-1}$ (PSR
1952+29). Therefore the increment of the instability must not be
less than $2\dot P/P$.

We investigate the instability rate nearby the light
 cylinder, because the centrifugal acceleration should be most
 efficient in this region. In Fig. \ref{lam} we show dependence of the increment on:
$\lambda_{x}$ nearby the LCS. The set of parameters is $P = 1s$,
$\gamma_b\sim 10^8$, $\lambda_x\approx\lambda$, $\lambda_{\phi} =
1000R_{lc}$ and it is supposed that $k_x<0$ and $U_{bx}>0$
(otherwise the resonance frequency is unphysical - negative). Here
$R_{lc}$ is the light cylinder radius. Such a choise of
$\lambda_{\phi}$ provides almost perpendicular (to the equatorial
plane) propagation of waves. One can see that the increment
reaches the value $\sim 10^{-2}$, which is more by many orders of
magnitude than typical values of $2\dot P/P$. The linear stage
will be very efficient and short in time strongly indicating the
non linear regime of the phenomenon. A need of non linear
saturation is seen also from the fact that $B_r$, which is
responsible for twisting of magnetic field lines, oscillates with
frequency $\omega_0$, due to this oscillation, $B_r$ not only will
lag behind the rotation, which is physically reasonable, but also
will advance it, therefore the need of non linear saturation of
the instability increment is essential. Therefore initially
created small perturbations will rapidly increase in time and
thanks to the instability process, it will extract energy from the
background flow into the energy of electrostatic waves.

\section{Summary} \label{sec:summary}
%
%
%
%

\begin{enumerate}
      \item Considering the relativistic plasma flow composed of
      the primary and secondary (beam) components, we have studied
      the role of the centrifugal acceleration in the curvature drift
      instability.

      \item Making the linear analysis of equations, we have
      derived the dispersion relation and a
      new mechanism of the parametric instability responsible for
      rotation energy pumping has been found.

      \item Considering low frequencies, which are responsible for twisting of
      magnetic field lines, an expression for the
      instability increment has been obtained.

      \item Studying dependence of increment on
      $\lambda_{x}$, it has been found that the instability
      was very efficient and increments were more than pulsar spin down rates by many order of
      magnitude indicating the need of the non linear consideration of
      the problem.

As we have seen, the analysis indicated the importance of the non
linear stage in dynamics of the instability, therefore it is
essential to study the same problem numerically by implementing a
special relativistic MHD code, which will comprise one more step
closer to the real scenario.

      \end{enumerate}

\section*{Acknowledgments}

The research was supported by the Georgian National Science
Foundation grant GNSF/ST06/4-096.

\appendix

\section{}

In this section we would like to show that the  sum:

\begin{equation}
\label{sumation} \Sigma_1(g_b,h)\equiv \sum_sJ_s(g_b)J_{-2s}(h),
\end{equation}
is finite.

In order to prove the convergence of (A.1) we  use the following
inequality:

\begin{equation}
\label{ineq1}
|J_{\nu}(x)|\leq\frac{1}{\Gamma(\nu+1)}\left(\frac{x}{2}\right)
^{\nu}e^{Im[x]}.
\end{equation}
Here: $\nu\geq-\frac{1}{2}$.

Then for the term in $\Sigma_1(g_b,h)$, one can write:

\begin{equation}
\label{ineq2}
|J_s(g_b)J_{2s}(h)|\leq\frac{1}{(s+1)!(2s+1)!}\left(\frac{g_b}{2}\right)^{s}\left(\frac{h}{2}\right)^{2s+1}.
\end{equation}
where we have used the well known equivalence:

\begin{equation}
\label{eqv} J_{-\nu}(x)=(-1)^{\nu}J_{\nu}(x).
\end{equation}

This condition shows that $C_{s}\equiv |J_s(g_b)J_{2s}(h)|\leq
U_{s}$, where:
\begin{equation}
\label{ineq3}
U_{s}=\frac{1}{(s+1)!(2s+1)!}\left(\frac{g_b}{2}\right)^{s}\left(\frac{h}{2}\right)^{2s+1}.
\end{equation}

Let us prove that the $U_{s}$ is convergent using the Dalamber
criterion, by introducing the following ratio:

\begin{equation}
\label{cond1}
\frac{U_{s+1}}{U_{s}}=\left(\frac{g_b}{2}\right)\left(\frac{h}{2}\right)^{2}\frac{1}{(s+2)(2s+2)(2s+3)}.
\end{equation}

It is obvious that one can find $s_{0}$ for which
$\frac{U_{s_0+1}}{U_{s_{0}}}\equiv q<1$. We see that the
expression in Eq. (\ref{cond1}) decreases with an increasing value
of $s$, i.e:

\begin{equation}
\label{cond2}
\frac{U_{s+1}}{U_{s}}<q=\left(\frac{g_b}{2}\right)\left(\frac{h}{2}\right)^{2}\frac{1}{(s_0+2)(2s_0+2)(2s_0+3)}<1,
\end{equation}
$$ s\geq s_{0}.$$

This means that the sequence  $U_{s}$ is convergent, and hence for
$s\geq 0$ the summation $\Sigma_1(g_b,h)$ is finite.

When considering the case $s<0$ and formally introducing a new
index $ m\equiv -s$,  one obtains an expression:

\begin{equation}
\label{ineq4} |J_m(g_b)J_{2m}(h)|,
\end{equation}
similar to a corresponding term in Eq.(\ref{ineq2}) for $s\geq 0$
and hence, the summation for negative values of $s$ is  also
convergent.

The proof for convergence of the second summation
$\Sigma_2(g_b,h)$ (see Eq. (\ref{sig2})) does not principally
differ from the one we have already considered and therefore we do
not show it here.

\end{document}